\documentclass[12pt, aps,amsmath,showpacs,superscriptaddress, pra]{revtex4-1}
\usepackage{dcolumn,bbm,amssymb,amsmath,ulem,indentfirst,amsthm,color,graphicx}
\begin{document}
\title{The partition function in the quantum-to-classical transition}

\author{Bingyu Cui}
\email{bycui@cuhk.edu.cn}
\affiliation{School of Science and Engineering, The Chinese University of Hong Kong, Shenzhen, Guangdong 518172, PR China}
\date{\today}

\begin{abstract}
\noindent In classical statistical mechanics, the partition function is defined in phase space. We extend this concept to quantum statistical mechanics by incorporating Bohmian trajectories. The quantum partition function formulated in phase space encompasses the ensemble of particle positions and momenta, together with a probability distribution that reflects the intrinsic uncertainty in determining particle locations. Within this framework, the proposed partition function smoothly reduces to the conventional expressions in the quantum and classical limits. The quantum-to-classical transition emerges naturally when the temperature exceeds a critical value, under a Gaussian form of the distribution, maintaining consistency between dynamical and statistical mechanical descriptions.
\end{abstract}


\maketitle

\section{Introduction}
Statistical mechanics bridges microscopic physical laws with macroscopic descriptions of nature. By employing statistical methods, it reveals emergent phenomena that may not be explicit in the underlying fundamental laws. Quantum systems, regardless of size, are described by microstates corresponding to solutions of the Schr\"{o}dinger equation, whereas classical systems are described by phase-space microstates, which specify particle positions and momenta. Although many concepts in quantum mechanics carry over to classical macrostates, the manner in which the two frameworks yield convergent results remains conceptually subtle and is tied to the longstanding problem of the quantum-to-classical transition \cite{Ehrenfest1927,Dirac1945,landau2013,Rosen1964,Zeh1970,Home1983,ballentine2013,Home1995,Ghose2002,ValiaAllori2002,Zurek2003,diosi2003,Mosna2005,Vitiello2005,schlosshauer2007,Allori2008,Home2009,herbert2011,Bondar2012,Sanz2012,joos2013,Nassar2013,Gomez2014,Lusanna2014,Ghose2015,Briggs2016,Budiyono2017,Oriols2016,Oriols2017,Demme2017,Chou2018,Konishi2022}.

In classical statistical mechanics, despite the deterministic nature of Newton’s laws, precise knowledge of every particle’s position and momentum is typically inaccessible, necessitating a statistical description in terms of ensembles and averages. In quantum mechanics, even individual observables—such as position or momentum—are intrinsically uncertain. Bohmian mechanics provides a complementary perspective: Particles have well-defined positions and momenta at all times, guided by the wavefunction, but these variables are hidden from direct observation. Observable quantities emerge as ensemble averages over the probability distribution given by the wavefunction’s squared amplitude \cite{Bohm1952,BOHM1966,BELL1966,holland1993,Drr2013}. Compared to standard quantum mechanics, Bohmian mechanics offers a clearer and more intuitive interpretation of experimental results in terms of particle trajectories \cite{Home1999,Guantes2004,Oriols2007,Das2019}.

The empirical probabilities computed in Bohmian mechanics agree with standard quantum mechanics when the ensemble of systems is distributed according to the quantum equilibrium hypothesis \cite{Durr1992}. Because Bohmian mechanics employs the same fundamental variables as classical mechanics—positions and momenta—it offers a natural phase-space framework for unifying statistical mechanics across quantum and classical regimes. Nevertheless, only a limited number of studies have explored quantum statistical mechanics within the Bohmian context  \cite{bohm2006undivided,Bohm1996,Valentini1991,Valentini1991B,Drr1992,Durr2005449,allori2020statistical,Valentini2020,bricmont2022making,Nikoli2024}, and fewer still have examined the quantum-to-classical transition from this viewpoint.

In this letter, we extend the notion of phase space to closed quantum systems using Bohmian trajectories and formulate a generalized partition function in the canonical ensemble. This partition function is expressed over positions and momenta and includes weight functions that represent deviations in localization, which—in quantum mechanics—are identified with probability distributions. We demonstrate applicability to both small and large systems.  Building on Ref. \cite{Cui2023}, where a protocol was introduced for recovering classical dynamics in the center-of-mass motion of large quantum systems (with the probability distribution becoming $\delta$-like as system size grows), we show that the proposed partition function captures a consistent quantum-to-classical transition within statistical mechanics. Since thermodynamical quantities are computed from the partition function, their expressions from the proposed formula also agree in quantum and classical limits.  We further derive a criterion of temperature to judge the classicality.

The structure of this letter is organized as follows: Section II reviews core principles of dynamics and statistical mechanics in classical and quantum frameworks. Section III presents the proposed partition function based on Bohmian trajectories, followed by the discussion of the quantum-to-classical crossover in Sec. IV. Section V illustrates the approach with a harmonic oscillator. Finally, we conclude in Sec. VI.

\section{Classical and quantum mechanics}
In this section, we briefly review and compare the general foundations of (non-relativistic) classical and quantum (statistical) mechanics for closed systems. \\

\subsection{Classical dynamics and statistical mechanics}
Classical statistical mechanics is based on a phase space formulation. For a closed system of $N$ classical particles with positions $\mathbf{x}_n$ and momenta $\mathbf{p}_n, n=1,...,N$, the precise microstate of the system is specified by a representative vector, $\vec{z}=(z_1,...,z_N)$, where $z_n=(\mathbf{x}_n,\mathbf{p}_n)$ is a point in the 6-dimensional phase space of the $n$th particle. Consider the movement of a system of classical particles of (same) mass $m$ subject to an external potential $V(\vec{x},t),\vec{x}=(\mathbf{x}_1,\mathbf{x}_2,...,\mathbf{x}_N)$. The movement of the n-th particle is governed by Newton's second law of motion:
\begin{equation}
    m\frac{d^2\mathbf{x}_n}{dt^2}=-\nabla_{\mathbf{x}_n} V(\vec{x},t).
    \label{eq:Newton}
\end{equation}
Another viewpoint of the classical dynamics from time $t_i$ to $t_f$ (and $t=t_f-t_i$ is the duration) is the Hamilton-Jacobi equation \cite{leech2012,goldstein1950,Oriols2019}:
\begin{equation}
    \frac{\partial S(\vec{x},t)}{\partial t}+H\left(\vec{x},\vec{p},t\right)=0,
    \label{eq:HamiltonJacobi}
\end{equation}
in which the Hamiltonian and the Hamilton principal function $S(\vec{x},t)$ are,
\begin{align}
\label{eq:Hamilton}
    &H(\vec{x},\vec{p},t)=\sum_{n=1}^N\frac{(\nabla_{\mathbf{x}_n}S)^2}{2m}+V(\vec{x},t),\\
    &S(\vec{x},t)=\int_{t_i}^{t_f}\left[\sum_{n=1}^N\frac{(\nabla_{\mathbf{x}_n}S)^2}{2m}\right]-V(\vec{x},t)d\tau,
    \label{eq:action}
\end{align}
Note that the momenta are given by $\mathbf{p}_n=\nabla_{\mathbf{x}_n}S$ and $\vec{p}=(\mathbf{p}_1,\mathbf{p}_2,...,\mathbf{p}_N)$.

To proceed in statistical mechanics, imagine an ensemble of identical classical systems, for which one can define the probability density $\rho(\vec{z},t)$ in the full 6$N$-dimensional phase space, which is equal to the fraction of systems located within an infinitesimal volume $d\Gamma$ surrounding the point $\vec{z}$. The infinitesimal volume is given by
\begin{equation}
    d\Gamma\equiv\prod_{n=1}^N\frac{d\mathbf{x}_nd\mathbf{p}_n}{(2\pi\hbar)^3}.
    \label{eq:diffvolume}
\end{equation}
The occurrence of the factor $2\pi\hbar$ in the definition of volume $\Gamma$ does not matter for any physical observable. Since particles just follow their respective trajectories in a statistical ensemble, the probability density satisfies the continuity equation
\begin{equation}
    \frac{\partial \rho}{\partial t}+\sum_{n=1}^N\nabla_{z_n}\cdot \mathbf{j}_n=0,
    \label{eq:manycontCM}
\end{equation}
with
\begin{equation}
    \mathbf{j}_n=\rho\left(\frac{\partial H}{\partial \mathbf{p}_n},-\frac{\partial H}{\partial \mathbf{x}_n}\right)
\end{equation}
the current in the 6-dimensional subspace. It then follows from Liouville's theorem that the probability density is constant along system trajectories in phase space and satisfies the Liouville equation, 
\begin{equation}
    \frac{\partial\rho}{\partial t}=\{H,\rho\},
\end{equation}
where $\{,\}$ is the Poisson bracket. The partition function is defined as the integration over phase space 
\begin{equation}
    Z_{cl}=\int e ^{-\beta H(\vec{z})}d\Gamma,
    \label{eq:classicalparti}
\end{equation}
where $\beta=(k_BT)^{-1}$ and $k_B$ is the Boltzmann constant.

\subsection{The Bohmian interpretation of quantum mechanics}
Unlike classical objects, the measurement of physical observables like the position and momentum of a quantum particle is nondeterministic. For a closed quantum system, rather than classical equations of motion, one instead looks at the evolution of the wavefunction, which obeys many-particle Schr\"{o}dinger equation:
\begin{equation}
    i\hbar\frac{\partial\Psi(\vec{x},t)}{\partial t}=\left(\sum_{n=1}^N-\frac{\hbar^2}{2m}\nabla_{\mathbf{x}_n}^2+V(\vec{x},t)\right)\Psi(\vec{x},t)
    \label{eq:manyQE}
\end{equation}
where we have assumed all particles carry the same (non-zero) mass $m$. According to Bohm \cite{Bohm1952}, the wavefunction serves as a guidance field to lead the movement of particles. To see this, writing the wavefunction in polar form $\Psi=R(\vec{x},t)\exp[iS(\vec{x},t)/\hbar]$ with $P(\vec{x},t)\equiv R^2=|\Psi|^2$ and substituting it into Eq. \eqref{eq:manyQE}, we obtain two coupled equations from imaginary and real parts,
\begin{align}
\label{eq:manycontQM}
    &\frac{\partial P}{\partial t}+\frac{1}{m}\sum_{n=1}^N\nabla_{\mathbf{x}_n}\cdot(P\nabla_{\mathbf{x}_n}S)=0,\\
    &\frac{\partial S}{\partial t}+\sum_{n=1}^N\frac{(\nabla_{\mathbf{x}_n}S)^2}{2m}+V(\vec{x},t)-\frac{\hbar^2}{2mR}\sum_{n=1}^N\nabla_{\mathbf{x}_n}^2R=0.
    \label{eq:manyHJ}
\end{align}
Upon defining the velocity of $n$th particle $\mathbf{v}_n=\nabla_{\mathbf{x}_n}S(\vec{x},t)/m$ similarly as the classical Hamilton-Jacobi formalism, Eq. \eqref{eq:manycontQM} can be viewed as the continuity equation of the probability density of representative points $\vec{x}$. Despite resembling the continuity equation \eqref{eq:manycontCM} in the classical case, Eq. \eqref{eq:manycontQM} is well defined even for a single quantum particle \footnote{In classical systems, we could still consider an ensemble of a single system without knowing its exact state very well. However, here we are distinguishing the randomness in classical and quantum dynamics.}. In quantum mechanics, the ensemble corresponds to an infinite repetition of identical experiments. The 2nd equation \eqref{eq:manyHJ}, known as the quantum Hamilton-Jacobi equation, describes the movement of particles guided by the regular mechanical potential $V(\vec{x},t)$. Compared to its classical counterpart Eq. \eqref{eq:HamiltonJacobi}, an additional term, known as the quantum-mechanical potential,
\begin{equation}
    Q(\vec{x},t)\equiv-\frac{\hbar^2}{2mR}\sum_{n=1}^N\nabla_{\mathbf{x}_n}^2R,
\end{equation}
also changes particle trajectories. This can be seen by taking the gradient of Eq. \eqref{eq:manyHJ}, which implies a Newton-like equation of motion,
\begin{equation}
    m\frac{d^2\mathbf{x}_n}{dt^2}=-\nabla_{\mathbf{x}_n} [V(\vec{x},t)+Q(\vec{x},t)].
    \label{eq:quantumNewton}
\end{equation}
The probability distribution $P(\vec{x},t)$ and the Hamilton-Jacobi function $S(\vec{x},t)$ are coupled and co-determine each other via $Q(\vec{x},t)$, distinguishing quantum and classical dynamics. Equation \eqref{eq:quantumNewton} is deterministic and defines unique quantum (Bohmian) trajectories given the initial condition \cite{Oriols2019}. To incorporate the nondeterministic behavior in dynamics, it is Bohm's idea that quantum randomness arises from the uncontrollable precise (initial) location of the particle \cite{Bohm1952}, which, by the Born rule, forming the distribution $P(\vec{x},t)$, indicating the probability density function of detecting particles at respective positions $\vec{x}$ at time $t$.

Further, it is evident from Eq. \eqref{eq:manyHJ} that the energy of the system is \cite{Bohm1952}
\begin{equation}
    E(\vec{x},t)=-\frac{\partial S}{\partial t}=\sum_{n=1}^N\frac{(\nabla_{\mathbf{x}_n}S)^2}{2m}+V(\vec{x},t)+Q(\vec{x},t).
    \label{eq:energy}
\end{equation}
It can be shown that the expected value of $E(\vec{x},t)$ agrees with the result of standard quantum mechanics (see Appendix A for details).

\subsection{Partition functions in quantum statistical physics}
In quantum statistical physics, for a system with a fixed number of particles in thermal equilibrium, the partition function within the canonical ensemble counts all possible energy levels $E_m$ (correspond to eigenstates of the Hamiltonian $\hat{H}$ that consists of position operators $\hat{\mathbf{x}}_n$ and momentum operators $\hat{\mathbf{p}}_n$), which can be expressed as \footnote{Equation \eqref{eq:quantpf} is expressed for pure states. In quantum statistical mechanics, it is common to work with mixed states. Expressing the partition function in terms of mixed states, or density matrices, amounts to choosing different representations, but it does not change final value of the partition function.  }
\begin{equation}
    Z_q=\sum_k e^{-\beta E_k}.
    \label{eq:quantpf}
\end{equation}
If we neglect terms of non-zero orders of $\hbar$ (arises from the commutation between position and momentum operators) so that the eigenenergy spectrum becomes continuous \cite{Wong2014}, the quantum canonical partition function Eq. \eqref{eq:quantpf} consequently reduces to the classical canonical partition function \eqref{eq:classicalparti}. An alternative approach to the partition function and computation of thermodynamic averages for quantum systems is the path integral \cite{feynman2010quantum,feynman1972statistical}. It can be shown that the classical partition function is also attained when the temperature is high \cite{mcquarrie2000statistical,reif2009fundamentals}.

\section{From dynamics to statistical mechanics}
None of the aforementioned formalisms of the quantum partition function provides a clear physical insight into the transition between quantum and classical statistical mechanics. Unifying in phase space, we bridge the connection in statistical mechanics between quantum and classical systems.

First of all, note that to solve Newton's equation of motion \eqref{eq:Newton} in classical mechanics, (initial) conditions specifying position and momentum are required; distinct initial conditions generate unique trajectories. In phase space, a point $(\vec{x}(t),\vec{p}(t))$ maps uniquely to $(\vec{x}(t_0),\vec{p}(t_0))$ at any time $t_0$. For an ensemble of positions and momenta \footnote{In classical statistical mechanics, it is impossible to access the information of particles' (initial) locations and momenta.}, the classical partition function Eq. \eqref{eq:classicalparti} integrates over the thermal equilibrium distribution at a chosen snapshot (e.g. $t=0$).

We extend this phase-space picture to quantum systems via Bohmian trajectories and define the unified partition function for spinless particles as 
\begin{equation}
    Z_u=\sum_{\{P\}}\int P(\vec{x},t;\vec{z}(t))e^{-\beta E(\vec{x},t;\vec{z}(t))}d\vec{x}(t)d\vec{p}(t)d\vec{x},
    \label{eq:unipf}
\end{equation}
where $\vec{z}(t) \equiv (\vec{x}(t), \vec{p}(t))$ denotes the phase-space point specifying the instantaneous trajectory configuration at time $t$, and  $P(\vec{x},t;\vec{z}(t))=|\Psi(\vec{x},t;\vec{z}(t))|^2$ is the (conditional) probability density. This notation does not introduce a new wavefunction; rather, $\Psi(\vec{x},t;\vec{z}(t))$ emphasizes that the same guiding wavefunction generates the trajectory passing through  $\vec{z}(t)$. In Bohmian mechanics, particles have definite positions $\vec{x}(t)$ and momenta $\vec{p}(t)$, but these “hidden variables” are inaccessible to the external observer; the ensemble distribution $P$ encodes this uncertainty and serves as the weight associated with the phase-space point $\vec{z}(t)$. The energy $E(\vec{x},t; \vec{z}(t))$ is the energy along the trajectory defined by $\vec{z}(t)$ and the Hamiltonian.

The sum over $\{P\}$ in Eq. \eqref{eq:unipf} indicates counting over all distinct and physically admissible (stationary) distributions consistent with (thermal) equilibrium. Nonstationary distributions would induce a time dependence in the partition function, which is unphysical for equilibrium systems. Consequently, only stationary distributions, i.e., energy eigenstates of the Hamiltonian, contribute. Integrating out the hidden variables $(\vec{x}(t),\vec{p}(t))$ and summing over stationary distributions reduces \eqref{eq:unipf} to the standard quantum partition function \eqref{eq:quantpf}. Thus, the proposed formulation is equivalent to conventional quantum statistical mechanics in thermal equilibrium.

\section{Quantum to classical crossover}
The unified partition function \eqref{eq:unipf} recovers conventional results in the quantum and classical domains and provides a phase-space paradigm for the quantum-to-classical transition. In particular, Ref. \cite{Cui2023} demonstrates that the classical trajectory is recovered for the center of mass of a large quantum system whenever its probability distribution becomes Gaussian (by the central limit theorem) with the width $\sim\mathcal{O}(1/N)$ and $N$ is the number of particles.

Consider the normal distribution (in 1D) of a single particle carrying mass $m$ at time $t=0$,
\begin{equation}
    P_G(x;x(0),p(0))=\frac{1}{\sqrt{2\pi}\sigma}\exp\left[-\frac{(x-x(0))^2}{2\sigma^2}\right],
    \label{eq:Pt0}
\end{equation}
where $\sigma$ is the width and $(x(0),p(0))$ refers to the initial condition of a trajectory. The corresponding partition function is
\begin{equation}
    Z_u=\int P_G(x;x(0),p(0))\exp\left[-\beta\left(\frac{p(0)^2}{2m}+V(x(0))+\frac{\hbar^2}{4m\sigma^2}-\frac{\hbar^2(x-x(0))^2}{8m\sigma^4}\right)\right]dx(0)dp(0)dx.
    \label{eq:gausspart}
\end{equation}
The last two terms in the exponent are the quantum potential associated with the Gaussian-shaped distribution. Such a form of partition could correspond to the center of mass of a quantum system in an ensemble of its replicas. Thus, the product $m\sigma^2$ remains fixed and finite. Following the same way as the transition to classical dynamics for the center of mass, when the number of particles (constituting one realization represented by their center of mass) is large enough, $\sigma\rightarrow0$, and $P_G\rightarrow\delta(x-x(0))$, Eq. \eqref{eq:gausspart} reduces to the classical partition function Eq. \eqref{eq:classicalparti} (up to some multiplicative factor). In this case the ``particle" picks up the unique initial position $x(0)$, at which the quantum force $\partial Q/\partial x$ (cf. Eq. \eqref{eq:quantumNewton}) vanishes, recovering the classical dynamics. The distribution $P$ and energy $E$ are also stationary in the classical limit $\sigma\rightarrow0$.

Convergence of the integral in Eq.  \eqref{eq:gausspart} requires the coefficient of $(x - x_0)^2$ in the exponent to be negative, yielding the lower bound 
\begin{equation}
    T>\frac{\hbar^2}{4m\sigma^2k_B}.
    \label{eq:lowboundT}
\end{equation}
It is remarkable that the corresponding shortest dispersion $\sigma\sim\sqrt{\hbar^2/4mk_BT}$ is comparable to the thermal de Broglie wavelength $\lambda\sim\sqrt{2\pi\hbar^2/mk_BT}$. This bound, derived for Gaussian distributions typical of the transition regime, quantifies when thermal broadening suppresses quantum corrections in the partition function. Although it is model-dependent, it provides a useful diagnostic: If $T$ falls below this threshold, quantum effects remain relevant and classical approximations fail.

To identify other possibilities of approaching the classical limit, it is sufficient to look at the marginal partition function at a given initial position and momentum, $(x(0),p(0))\equiv(x_0,p_0)$,
\begin{equation}
    Z_{(x_0,p_0)}=\sum_{\{P\}}\int Pe^{-\beta E}dx,
    \label{eq:margin}
\end{equation}
whose time derivative is
\begin{equation}
    \frac{d Z_{(x_0,p_0)}}{dt}=\sum_{\{P\}}\int\left(\frac{d P}{d t}+P\frac{d E}{d t}\right)e^{-\beta E}dx.
\end{equation}
The first term in the bracket on the right is the time variation in the probability distribution. Even for a sharp normal distribution ($\sigma$ is non-zero), it is still possible that the particle's movement deviates from classical trajectories, experiencing a non-vanishing quantum force. Such a force might cause a change in the energy, as is shown in the second term in the bracket. 
If the temperature is high enough, or the particle energy $E$ is much smaller than the thermal energy $k_BT$ so that the spatial variation of $E$ in the exponent in Eq. \eqref{eq:margin} can be neglected, the (marginal) partition function becomes time-independent and reduces to the classical form.

We also note that thermodynamic observables follow from \eqref{eq:unipf} in the usual way. For example, the average energy is
\begin{align}
    \langle E\rangle&=-\frac{\partial}{\partial\beta}\log Z_u\\
    &=\frac{1}{Z_u}\sum_{\{P\}}\int P(\vec{x},t;\vec{z}(t))E(\vec{x},t;\vec{z}(t))e^{-\beta E(\vec{x},t;\vec{z}(t))}d\vec{x}(t)d\vec{p}(t)d\vec{x},
    \label{eq:avgE}
\end{align}
In the pure quantum regime, integrating out hidden variables and retaining only stationary distributions (energy eigenstates) $\{\phi_k(\vec{x})\}$ gives $Z_u\rightarrow Z_q$ and the numerator reduces to
\begin{align}
    \sum_{\{P\}}\int P(\vec{x},t;\vec{z}(t))E(\vec{x},t;\vec{z}(t))e^{-\beta E(\vec{x},t;\vec{z}(t))}d\vec{x}(t)d\vec{p}(t)d\vec{x}\rightarrow\sum_k E_ke^{-\beta E_k}\int|\phi_k(\vec{x})|^2d\vec{x},
\end{align}
which recovers the standard expression for $\langle E\rangle$ in standard quantum statistical mechanics. On the other hand, in the classical limit, substituting $P$ (or $P_G$)$\rightarrow\delta(\vec{x}-\vec{x}(0))$ into Eq. \eqref{eq:avgE} yields
\begin{align}
    \langle E\rangle\rightarrow\langle E\rangle_C+\frac{\hbar^2}{4m\sigma^2},\
\end{align}
with
\begin{align}
    \langle E\rangle_C=\frac{\int H(\vec{z})\exp\left[-\beta H(\vec{z})\right]d\Gamma}{\int \exp\left[-\beta H(\vec{z})\right]d\Gamma},
\end{align}
the classical average energy. The additive constant $\hbar^2/(4m\sigma^2)$ shifts the absolute energy; it does not affect thermodynamic derivatives such as heat capacity, so all standard thermodynamic relations are preserved.

\section{Single harmonic oscillator}
In one example, we calculate the partition function for a single harmonic oscillator with mass $m$ and frequency $\omega$ in 1D. Another example of a free particle is presented in Appendix B. Note that, because of the many-body problem, it is difficult to directly apply the Eq. \eqref{eq:unipf} to large quantum systems. However, the formalism stated above might provide new insight in other situations. In Appendix C, we use the proposed partition function to elucidate the quantum-to-classical crossover of the thermal reservoir in the Caldeira-Leggett system \cite{Caldeira1983}. Starting from the wavepacket, the time evolution of the wavefunction $\psi$ is
\begin{align}
    \psi(x, t; x(0),p(0),\sigma^2)&=A\exp\left[-\alpha(t)(x-q(t))^2+\frac{i}{\hbar}p(t)(x-q(t))+\frac{i}{\hbar}\gamma(t)\right],
    \label{eq:ho}
\end{align}
with
\begin{subequations}
\begin{align}
    \alpha(t)&=\frac{m\omega}{\hbar}\frac{\hbar\cos(\omega t)+i2\sigma^2m\omega\sin(\omega t)}{i2\hbar\sin(\omega t)+4\sigma^2m\omega\cos(\omega t)},\\
    q(t)&=x(0)\cos(\omega t)+\frac{p(0)}{m\omega}\sin(\omega t),\\
    p(t)&=p(0)\cos(\omega t)-m\omega x(0)\sin(\omega t),\notag\\
    \gamma(t)&=i\frac{\hbar^2a}{m\omega}\ln\left[\frac{i\sin\phi\sin(\omega t)+\cos\phi\cos(\omega t)}{\cos\phi}\right]\\
    &+\left(\frac{p(0)^2}{2m}-\frac{m\omega^2x(0)^2}{2}\right)\frac{\sin(2\omega t)}{2\omega}+\frac{p(0)x(0)\cos(2\omega t)}{2}+\gamma
\end{align}
\end{subequations}
where $A$ is a normalization constant; $q(t)$ and $p(t)$ are the position and momentum of the center of the wavepacket, respectively; the real constant $\gamma$ in $\gamma(t)$ does not lead to anything physical and will be ignored. The probability distribution remains the Gaussian shape, and 
\begin{align}
    \text{Re}[\alpha(t)]&=\frac{1}{4\sigma^2[\cos^2(\omega t)+\tan^2\phi\sin^2(\omega t)]},
\end{align}
with
\begin{align}
    \tan\phi&\equiv\frac{\hbar}{2\sigma^2m\omega}.
\end{align}

\begin{figure}[!tp]
\includegraphics[width=0.8\textwidth]{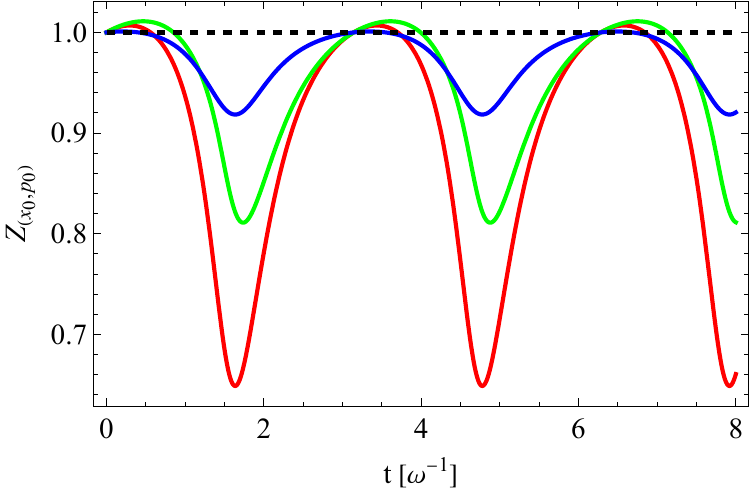}
\caption{Time evolution of the marginal partition function $Z_{(x_0,p_0)}$ in Eq. \eqref{eq:margin} for a quantum harmonic oscillator with mass $m$ and frequency $\omega$. The initial state is the Gaussian wavepacket in Eq. \eqref{eq:ho} with specified $(x(0),p(0))\equiv(x_0,p_0)$. Curves show $\sigma=0.45,k_BT=2$ (red); $\sigma=0.45,k_BT=5$ (blue) and $\sigma=0.65,k_BT=2$ (green). All curves are normalized to $Z_{(x_0,p_0)}=1$ at $t=0$ (black dashed line). The displacement is measured in units of $x_0$; energy in units of $m\omega^2x_0^2$ and time in units of $\omega^{-1}$.}
\label{fig:1}
\end{figure}

The energy of the particle is 
\begin{align}
    E(x,t; x(0),p(0))=\frac{p(t)^2}{2m}+\frac{m\omega^2q(t)^2}{2}+Q,
\end{align}
where
\begin{align}
    Q(x,t;x(0),p(0))&=-\frac{2\hbar^2a^2}{m}\frac{a^2\alpha_0^2-(a^2-\alpha_0^2)^2\sin^2(\omega t)\cos^2(\omega t)}{[\alpha_0^2\sin^2(\omega t)+a^2\cos^2(\omega t)]^2}(x-q(t))^2\notag\\
    &-\frac{2a\hbar p(t)}{m}\frac{(a^2-\alpha_0^2)\sin(\omega t)\cos(\omega t)}{\alpha_0^2\sin^2(\omega t)+a^2\cos^2(\omega t)}(x-q(t))\notag\\
    &+\frac{\hbar^2}{m}\frac{a^2\alpha_0}{\alpha_0^2\sin^2(\omega t)+a^2\cos^2(\omega t)}.
\end{align}
is the quantum potential with $a=m\omega/(2\hbar)$ and $\alpha_0=1/(4\sigma^2)$. In Fig. \ref{fig:1}, we show the marginal partition function at a given $(x(0),p(0))\equiv(x_0,p_0)$. Higher temperature suppresses oscillations in $Z_{(x_0, p_0)}$, while a smaller $\sigma$ enhances sensitivity to quantum forces away from the packet center, which is because even if it is rare, the particle would receive a larger quantum force from a sharper wavepacket when it does not sit at the center of the wavefunction.

\section{Conclusion}
In this letter, mimicking the idea of quantum trajectories embedded in phase space, we have generalized the partition function for closed quantum systems in the canonical ensemble. The stationary part of the partition function coincides with the conventional quantum expression and reduces to the classical partition function when the probability distribution sharpens to a delta-like form, consistent with the mechanism of classicalization seen in dynamics (e.g., center-of-mass motion in large systems) \cite{Cui2023}. It is thus not sufficient to consider only the formal limits $\hbar\rightarrow0$ or $T\rightarrow\infty$; rather, classical behavior emerges when the distribution becomes a narrow wavepacket at sufficiently high temperature, corresponding to a small thermal de Broglie wavelength and negligible interference. Since thermodynamic observables follow from the partition function via standard derivatives, and $Z_u$ reduces to $Z_q$ and $Z_{cl}$ in the respective limits, all conventional thermodynamic relations are recovered. While spin and indistinguishability are not treated here explicitly, both are encoded in the probability density and can be incorporated straightforwardly for bosons/fermions, even in relativistic generalizations  \cite{Ghose1994,Ghose2001,Drr2004,Struyve2010}.

For general quantum systems in thermal equilibrium, the stationary partition function admits only Hamiltonian eigenstates, in line with the spirit of the eigenstate thermalization hypothesis (ETH) \cite{Srednicki1994,Srednicki1999,Deutsch1991}. Non-ergodic systems - such as integrable models \cite{Kinoshita2006,Gring2012,Mori2018} or systems with quenched disorder \cite{Basko2006,Nandkishore2015} - violate ETH. The proposed partition function could provide a complementary lens on their thermally averaged observables. Furthermore, realistic thermodynamic systems are open, a comprehensive extension of this framework to open systems is a promising direction for future work.

\begin{appendix}
\section{The time evolution of the energy}
The energy (cf. Eq. \eqref{eq:energy} in the main text) of a quantum particle may be written in terms of the wavefunction and its complex conjugate:
\begin{align}
    E(\vec{x},t)&=\frac{i\hbar}{2}\frac{\Psi^*\dot{\Psi}-\Psi\dot{\Psi}^*}{|\Psi|^2}.
\end{align}
where $\dot{\Psi}$ is the time derivative of $\Psi$.
The mean particle energy is the average with the weighting function, $P=|\Psi|^2$,
\begin{align}
    \langle H\rangle&=\int PE(\vec{x},t)d\vec{x}\notag\\
    &=\frac{i\hbar}{2}\int (\Psi^*\dot{\Psi}-\Psi\dot{\Psi}^*)d\vec{x}.
    \label{eq:aveH}
\end{align}
Same as in the standard quantum mechanics, the mean energy is constant in time, 
\begin{equation}
    \frac{d\langle H\rangle}{dt}=\int\left(\frac{d P}{d t}+P\frac{d E}{d t}\right)d\vec{x}=0.
\end{equation}
To see this, we can choose a representation of energy eigenstates $\{\phi_k(\vec{x})\}$ with $E_k$ the corresponding eigenenergies. The wavefunction $\Psi$ might be decomposed as 
\begin{equation}
    \Psi(\vec{x},t)=\sum_kc_ke^{-iE_kt/\hbar}\phi_k(\vec{x})
    \label{eq:psiphi}
\end{equation}
with 
\begin{equation}
    c_k\equiv\int \phi_k^*\Psi(x,t=0)d\vec{x}.
\end{equation}
Substituting Eq. \eqref{eq:psiphi} into Eq. \eqref{eq:aveH} and making use of the orthonormality of $\{\phi_k(\vec{x})\}$, i.e.,
\begin{equation}
    \int \phi_k^*(\vec{x})\phi_s(\vec{x})d\vec{x}=\delta_{ks},
\end{equation}
it yields that the average energy is invariant in time,
\begin{equation}
    \langle H\rangle=\sum_k|c_k|^2E_k.
\end{equation}

\section{Free particle}

In the 2nd example, we consider a free particle carrying mass $m$. The time evolution of the wavefunction is 
\begin{subequations}
\begin{align}
    \psi(x, t; x(0),p(0),\sigma^2)&=A\exp\left[-\alpha(t)(x-q(t))^2+\frac{i}{\hbar}p(t)(x-q(t))+\frac{i}{\hbar}\gamma(t)\right],\\
    \alpha(t)&=\frac{1}{4\sigma^2\left(1+i\frac{2\hbar\alpha_0t}{m}\right)},\\
    q(t)&=\frac{p(0)t}{m}+x(0),\\
    p(t)&=p(0),\\
    \gamma(t)&=-\frac{p(0)^2t}{2m}+\frac{i\hbar}{2}\ln\left(1+\frac{2i\hbar t}{4\sigma^2m}\right)+\gamma,
\end{align}
\end{subequations}
where $A$ is a normalization constant, $q(t),p(t)/m$ are the position and velocity of the center of the wavepacket, respectively. The initial spread is $\sigma$, and the real constant $\gamma$ in $\gamma(t)$ can be safely ignored because it gives nothing useful. The corresponding probability distribution and energy are 
\begin{equation}
    P(x,t;x(0),p(0))=\frac{1}{\sqrt{2\pi(\sigma^2+\frac{\hbar^2t^2}{4m^2\sigma^2})}}\exp\left[-\frac{2(x-q(t))^2}{4\sigma^2+\frac{\hbar^2t^2}{m^2\sigma^2}}\right]
\end{equation}
and

\begin{align}
    E(x,t;x(0),p(0))&=\frac{p(t)^2}{2m}-\frac{2\hbar^2}{m}\frac{\alpha_0^2\left[1-\left(\frac{2\hbar\alpha_0t}{m}\right)^2\right]^2}{\left[1+\left(\frac{2\hbar\alpha_0t}{m}\right)^2\right]^2}(x-q(t))^2\notag\\
    &+\frac{2\hbar\alpha_0p(t)}{m}\frac{\frac{2\hbar\alpha_0t}{m}}{1+\left(\frac{2\hbar\alpha_0t}{m}\right)^2}(x-q(t))+\frac{\hbar^2}{m}\frac{\alpha_0}{1+\left(\frac{2\hbar\alpha_0t}{m}\right)^2}.
\end{align}
with $\alpha_0=1/(4\sigma^2)$. Note that the initial energy at $t=0$,
\begin{equation}
    E(x,0;x(0),p(0))=\frac{p(0)^2}{2m}-\frac{2\hbar^2\alpha_0^2}{m}(x-x(0))^2+\frac{\hbar^2\alpha_0}{m},
\end{equation}
is never reached again. In other words, unless the temperature is high enough, the information about the initial energy and thus the (marginal) partition function is washed out by the evolution of the wavefunction. 

\section{The quantum-to-classical transition of a heat bath}
In this section, we consider the Caldeira-Leggett (CL) particle-bath model consisting of a tagged particle immersed in a thermal bath of harmonic oscillators. For convenience, we focus on 1D. The Hamiltonian is
\begin{equation}
    H_{CL}=H_S+H_B,
    \label{eq:HCL}
\end{equation}
with
\begin{equation}
    H_S=\frac{p^2}{2m}+V(q),
\end{equation}
the Hamiltonian of the particle at position $q$ and momentum $p$, carrying mass $m$, and
\begin{equation}
    H_B=\frac{1}{2}\sum_{\alpha=1}^N\left(\frac{P^2_\alpha}{m_\alpha}+m_\alpha\omega_\alpha^2\left(X_\alpha-\frac{c_\alpha q}{\omega_\alpha^2}\right)^2\right)
\end{equation}
the Hamiltonian describing a collection of $N$ harmonic oscillators at positions $X_\alpha$ and momenta $P_\alpha$, carrying masses $m_\alpha$ and oscillatory frequencies $\omega_\alpha,
\alpha=1,2,...,N$. The coupling is linear in the particle coordinate $q$, where $c_\alpha$ is the coupling strength between the tagged particle and the $\alpha$th bath oscillator. The tagged particle and bath oscillators in the CL system can be quantum or classical \cite{Caldeira1983,Weiss2011,Zwanzig2001}. The direct implication and variation of the CL model, for example, when the tagged particle and/or bath oscillators carry charges, has received extensive studies \cite{Gupta2011,Cui2018, Grabert2018,Alpomishev2024,Gamba2024,Gamba2025}. The motion of the tagged particle is elucidated via the generalized Langevin equation \cite{Cui2018,Gamba2025}
\begin{equation}
    \frac{dp}{dt}=-V'(q)-\int_0^tv(s)\frac{P(t-s)}{m}ds+F_p(t),
\end{equation}
where
\begin{equation}
    F_p(t)=\sum_{\alpha=1}^N\left\{m_\alpha c_\alpha\left[X_\alpha(0)-\frac{c_\alpha Q(0)}{\omega_\alpha^2}\right]\cos(\omega_\alpha t)+c_\alpha P_\alpha(0)\frac{\sin(\omega_\alpha t)}{\omega_\alpha}\right\},
\end{equation}
is the noise or stochastic force and
\begin{equation}
    \nu(t)=\sum_{\alpha=1}^N\frac{c_\alpha^2}{\omega_\alpha^2}\cos(\omega_\alpha t)
\end{equation}
the memory kernel for the friction. Note that, in the quantum case, observables above are operators. If we assume the CL system is classical and at $t=0$ bath oscillators satisfy the Boltzmann distribution $\sim\exp(-H_B/k_BT)$. The average of a quantity $A$ associated with tagged particle is expressed as \cite{Zwanzig2001,Cui2018}
\begin{equation}
    \langle A\rangle=\frac{\int A\exp\left(-\frac{H_B}{k_BT}\right)d\vec{X}(0)d\vec{P}(0)}{Z_B},
\end{equation}
where $Z_B$ is the canonical partition function
\begin{equation}
    Z_B=\int \exp\left(-\frac{H_B}{k_BT}\right)d\vec{X}(0)d\vec{P}(0)
    \label{eq:ZB}
\end{equation}
and $\vec{X}(0)=\{X_1(0),X_2(0),...,X_N(0)\}, \vec{P}(0)=\{P_1(0),P_2(0),...,P_N(0)\}$. These results can be used to derive, for example, the fluctuation-dissipation relation responsible for the diffusion behavior of the tagged particle \cite{Gamba2024}.\newline
Now we relax the presumption that bath oscillators are classical and instead, we suppose the thermal bath is in the quantum-to-classical crossover. We further assume all oscillators follow the same (normal) distribution when $t=0$, cf. Eq. \eqref{eq:Pt0} in the main text, at which the total wavefunction is separable into particle and thermal baths. The canonical partition function $Z_B$ for the thermal reservoir takes the form
\begin{equation}
    Z_B'=\int P_Be^{-\beta E_B}d\vec{X}(0)d\vec{P}(0)d\vec{X},
\end{equation}
with
\begin{align}
    P_B&=\left[\frac{1}{\sqrt{2\pi}\sigma}\right]^N\exp\left[-\sum_{\alpha=1}^N\frac{(X_\alpha-X_\alpha(0))^2}{2\sigma^2}\right],\\
    E_B&=\sum_{\alpha=1}^N\left[\frac{P_\alpha^2(0)}{2m_\alpha}+\frac{m_\alpha\omega_\alpha^2}{2}(X_\alpha(0)-\frac{c_\alpha q(0)}{\omega_\alpha^2})^2+\frac{\hbar^2}{4m_\alpha\sigma^2}-\frac{\hbar^2(X_\alpha-X_\alpha(0))^2}{8m_\alpha\sigma^4}\right],
\end{align}
where $\vec{X}=\{X_1,X_2,...,X_N\}$. Integrating out $\vec{X}$, we obtain
\begin{equation}
    Z'_B=Z_B\cdot \prod_{\alpha=1}^N\left[2\pi \left(1-\frac{\beta\hbar^2}{4m_\alpha\sigma^2}\right)^{-1/2}\exp\left(-\frac{\beta\hbar^2}{4m_\alpha\sigma^2}\right)\right].
    \label{eq:ZBoverX}
\end{equation}
Thus, as long as the partition function \eqref{eq:ZBoverX} is well defined, average quantities take the same form as in the classical thermal baths. For $m_\alpha=m_0$ and large $N$,
\begin{equation}
    Z_B'\approx e^{-\frac{N\beta\hbar^2}{4m_0\sigma^2}}Z_B,
\end{equation}
which is valid if
\begin{equation}
    1-\frac{\beta\hbar^2}{4m_0\sigma^2}>0.
    \label{eq:LTbound}
\end{equation}
This reproduces the temperature criterion Eq. \eqref{eq:lowboundT} in the main text, and clarifies that, below this threshold, quantum effects in the bath cannot be neglected.

\end{appendix}

\section*{Author contributions}
B. C. is the unique author of this manuscript.

\section*{Acknowledgements}
The author wishes to thank Debankur Bhattacharyya, Henning Kirchberg and Zhen Tao for useful comments. This work was financially supported by the National Natural Science Foundation of China (No. 12404232), start-up funding from the Chinese University of Hong Kong, Shenzhen (No. UDF01003468) and the Shenzhen city “Pengcheng Peacock” Talent Program.

\section*{Statements and Declarations}

\subsection*{Conflict of interest}
The author declares no conflict of interest.

\section*{Data availability}
The data that support the findings of this study are available within the article.

\section*{Declaration of generative AI and AI-assisted technologies in the writing process}
During the preparation of this work, the author used GPT-5 to improve the language and readability of the main text. After using this tool, the author reviewed and edited the content as needed and takes full responsibility for the content of the publication.

\bibliography{reference}

\end{document}